\documentclass{aastex}

\usepackage{emulateapj5}

\newcommand{\xte}{{\it RXTE}}
\newcommand{\eps}{{\rm erg\,s^{-1}}}

\newcommand{\cts}{{\rm count\,s^{-1}}}
\newcommand{\kms}{{\rm km\,s^{-1}}}
\newcommand{\srco}{HD~154791}
\newcommand{\srcx}{4U~1700+24}
\newcommand{\asmend}{2002 April 18}
\newcommand{\asmendmjd}{52382}

\newcommand{\pall}{383}

\slugcomment{To appear in the Astrophysical Journal (2002)}

\shortauthors{Galloway et al.}
\shorttitle{}

\begin{document}


\title{
Correlated radial velocity and X-ray variations in \srco/\srcx\\
}


\author{Duncan K. Galloway}
\affil{Center for Space Research,
  Massachusetts Institute of Technology, \\Cambridge,MA 02139}
\email{duncan@space.mit.edu}
\and
\author{J. L. Sokoloski and Scott J. Kenyon}
\affil{Smithsonian Astrophysical Observatory, 60 Garden St., Cambridge, MA 02138}




\begin{abstract}
We present evidence for approximately 400-d variations in
the radial velocity of \srco\ (V934 Her), the suggested optical
counterpart of \srcx.  The variations are correlated with the previously
reported $\approx400$~d variations in the X-ray flux of \srcx, which
supports the association of these two objects,
as well as the identification of this system as the second known X-ray
binary in which a neutron star accretes from the wind of a red giant.
The \srco\ radial velocity variations 
can be fit with an eccentric orbit with period $404\pm3$~d, amplitude
$K=0.75\pm0.12\ \kms$ and eccentricity $e=0.26\pm0.15$. There are also
indications of variations on longer time scales $\ga2000$~d.  We have
re-examined all available ASM data following an unusually large X-ray
outburst in 1997--98, and confirm that the 1-d averaged 2--10~keV X-ray
flux from \srcx\ is modulated with a period of $400 \pm 20$~d.
The mean profile of the persistent X-ray variations was approximately
sinusoidal, with an amplitude of $0.108 \pm 0.012$~ASM~$\cts$ (corresponding to
31\% rms).
The epoch of X-ray maximum was approximately 40~d after the time of
periastron according to the eccentric orbital fit.
If the 400-d oscillations
from \srco/\srcx\ are due to orbital motion, then the system parameters
are probably close to those of
the only other neutron-star symbiotic-like binary, GX~1+4. We discuss the
similarities and differences between these two systems.
\end{abstract}


\keywords{binaries: symbiotic --- X-rays: binaries --- stars: neutron}


\section{Introduction}

The X-ray source \srcx\ (also known as 2A 1704+241; $l=45\fdg15$,
$b=32\fdg99$) was first discovered in {\it Ariel V}\/ scans for
high-latitude X-ray sources \cite[]{2acat}, and by the $Uhuru$ (SAS A)
X-ray observatory \cite[]{4ucat}.  It
has
a typical flux of (1--$10)\times10^{-11}\ {\rm erg\,cm^{-2}\,s^{-1}}$
\cite[2--10~keV, equivalent to 0.5--5~mCrab as measured by a range of
X-ray missions;][]{masetti01}, and experiences occasional outbursts.
During a 100-d X-ray high state in 1997--98,
\srcx\, was observed by the {\it Rossi X-ray Timing Explorer}\/
\cite[\xte;][]{xte96}.   This high state was the larger of
two known outbursts of this system, and the X-ray flux reached
a peak of 40~mCrab.
The X-ray spectrum of \srcx\, is similar to that of other accreting
neutron stars, and generally requires a blackbody with a temperature
of 0.9--1.3~keV plus a hard component
at higher energies for an acceptable fit.  
The X-ray spectral hardness and the lack of UV continuum imply that
the compact object is almost certainly not a white dwarf
\cite[]{garcia83}. A black hole companion is also unlikely since the thermal
X-ray component in black hole candidates is typically an order of
magnitude lower in temperature than that measured in \srcx. 
However, no rapid  quasi-periodic or coherent X-ray  
variations have been confirmed \cite[]{masetti01}, which is
rather unusual for an accreting neutron star.

Based on {\it Einstein}\/ position measurements, \cite{garcia83} proposed
the association of \srcx\ with the 8th magnitude M2 III star \srco.  In a
recent re-analysis of {\it ROSAT}\/ HRI data, however, \cite{mg01} found
only a 10\% chance that the two sources are associated, but did not
propose an alternative candidate.  Previous optical spectroscopy ruled out
radial velocity variations with amplitudes greater than $5\ \kms$ on time
scales of 40 min to $\sim 1$~yr \cite[]{garcia83}, strongly suggesting
that either the orbital period is significantly longer than one year, or
that the system is viewed very close to face-on (inclination angle
$i\approx0$).

\srco\ may be only the second symbiotic-like binary known to
contain a neutron star, after the well-studied M-giant/X-ray pulsar
system V2116~Oph/GX~1+4.  However, these systems are quite different in
X-rays (Table \ref{comptable}).  Furthermore, the optical spectrum of V2116~Oph
exhibits extremely strong, variable emission lines, probably powered by UV
photons originating from an accretion disk \cite[]{chak97:opt}.  In
contrast, \srco\ has much weaker line emission.  In fact, optically
the spectrum is very close to that of an isolated M giant
\cite[e.g.][]{garcia83}.

Here we describe optical and X-ray observations which go some way towards
resolving the uncertainties surrounding this system.  We compare the
properties of \srco/\srcx\ with those of V2116~Oph/GX~1+4, and discuss
what may be inferred about the former in light of our results.

\centerline{\epsfxsize=8.5cm\epsfbox{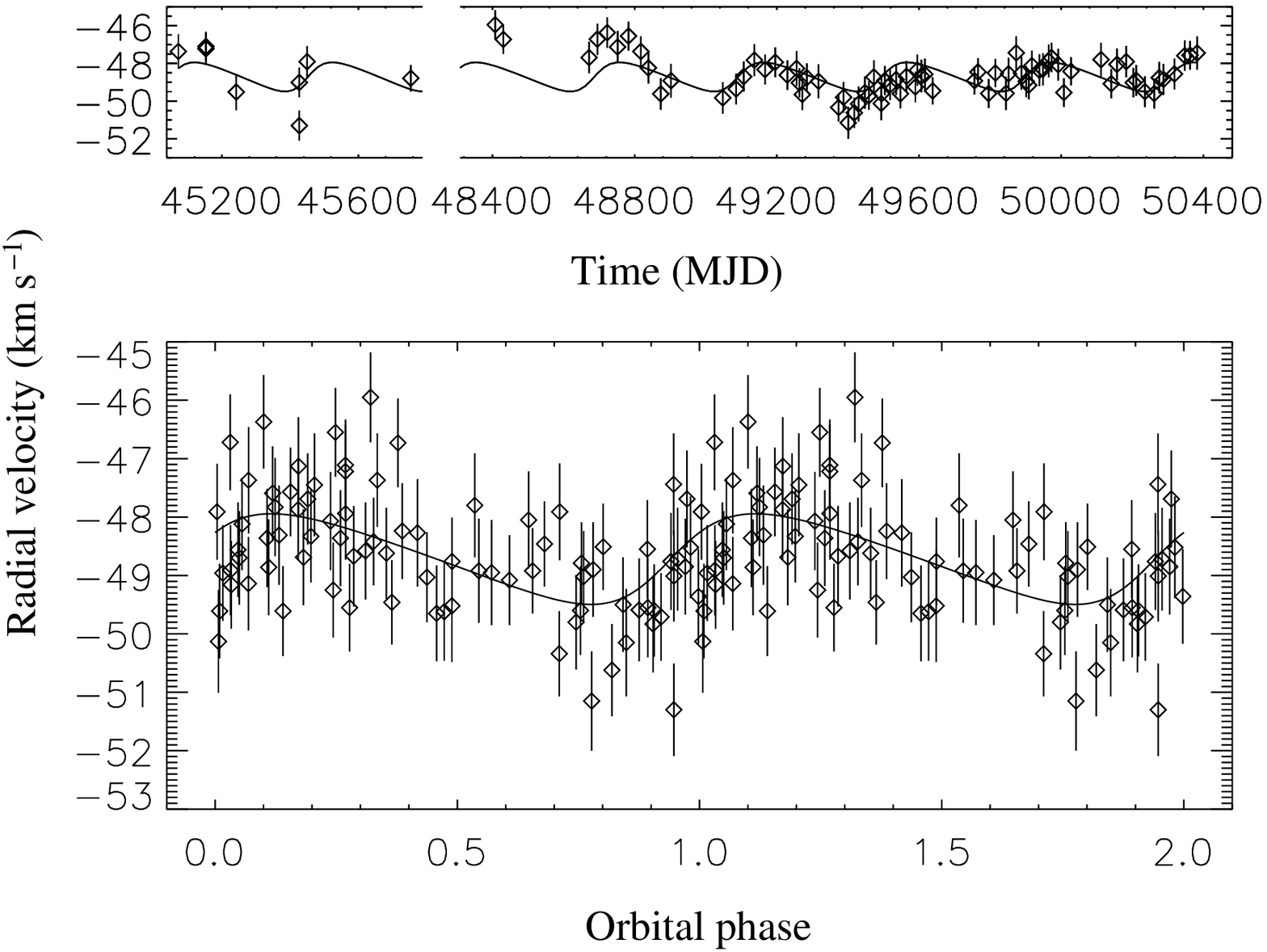}}
  \figcaption{Radial velocity measurements of \srco\ and
the corresponding $404\pm3$~d orbital solution. The top panel shows the
radial velocity measurements versus time and the estimated $1\sigma$
uncertainties.  The bottom panel shows the measurements folded on the
orbital period.  In both panels the eccentric orbital model is overplotted
as a solid line (see Table \ref{orbtbl}).
 \label{orbit} }
\bigskip

\section{Observations}

\subsection{Radial velocity variations} \label{sec:radvel}

Echelle spectra of \srco\ were obtained using intensified Reticon
detectors on the 1.5-m Tillinghast Reflector at the Whipple Observatory on
Mt. Hopkins, Arizona, and also the 1.5-m Wyeth reflector at Oak Ridge
Observatory, Massachusetts. The observations were made between 1982 April
12 and 1996 October 26, with the 18 observations before 1992 October 7
taken at Mt. Hopkins, and those since taken at Oak Ridge.  The instruments
are described in detail by \cite{latham92}. The spectra cover a 44~\AA\
bandpass centered around 5200~\AA, with a resolution of approximately $8.5\
\kms$. Each observation consisted of a 10--20 minute integration of \srco,
with 90~s Th-Ar comparison scans before and after.  The positions of
around 30 strong spectral lines in each calibration scan were fit with a
fifth-order polynomial, and the pre- and post-observation scans were then
combined to calibrate the wavelength scale for each spectrum. Typical
residuals to the polynomial fit were 0.5~pixel, corresponding to 0.01~\AA,
or $0.5\ \kms$. Approximately 25 lines were common to the entire set of
wavelength solutions.  The radial velocity was measured by
cross-correlating each observed spectrum with an M-star template spectrum
\cite[e.g.][]{td79} and measuring the mean offset.  Quadratic
least-squares fits to the peak in the cross-correlation function give a
radial velocity accurate to better than $1\ \kms$.

The radial velocity measurements show significant variations of
approximately $1\ \kms$ about the mean systemic velocity of $-48.7\pm0.1\
\kms$ (Figure \ref{orbit}). We searched the data for periodic signals in
the range 50--1000~d by calculating the Lomb-Scargle periodogram
\cite[e.g.][]{nr}, and found a marginally significant peak near a period
of 410~d (Lomb-normalized power 10.5; estimated significance from
bootstrap simulations, $3.3\sigma$).  Fitting an elliptical orbit to the
data gave a solution with orbital period $404\pm3$~d and radial velocity
amplitude $K=0.75\pm0.12\ \kms$ (Table \ref{orbtbl}; solid lines in Figure
\ref{orbit}).  The reduced-$\chi^2$ ($\chi^2_\nu$) for the fit was 1.26,
indicating a marginally acceptable fit. While the best-fit eccentricity
was 0.26, the deviations from circularity were significant at less than
the $2\sigma$ level.

There was also evidence for variations in the radial velocity on
longer time scales.  We calculated $\chi^2_\nu$ for an eccentric orbit
as a function of trial period for periods up to 4000~d.
While the 404~d signal gave rise to a $\chi^2_\nu$ value close to 1, lower
values were achieved at substantially longer periods around 1930~d
($\chi^2_\nu=1.04$), 2370~d (1.02) and 3700~d (0.91).  
The irregular spacing of the measurements did not allow us to distinguish
clearly between these periodicities; eccentric orbital solutions with any
one of the three periods can account for the long-term radial velocity
variations.  The observations do not cover more than one full cycle for
these longer-period signals, so more observations are needed to
determine whether they are true periodic variations or aperiodic trends
related to irregular red-giant variability.

\subsection{RXTE/ASM observations} \label{sec:xte}

\cite{masetti01} noted tentative evidence for an approximately
$400$~d periodicity in \xte\/ All-Sky Monitor (ASM) data.  To
investigate the 400-d periodicity, we obtained 1-d averaged ASM
measurements of \srcx\ between 1996 January 5 and
\asmend\ (MJD 50087--\asmendmjd) from the ASM WWW site at
{\tt http://xte.mit.edu}.  The mean quiescent ASM count rate was $0.2\
\cts$.  During the flare observed between 1997 September--December
(MJD 50700--50800), the 1-d averaged rate reached $2.5\ \cts$ (Figure
\ref{asm}, top panel).

To avoid a biased period measurement due to the presence of the 1997--98
outburst, we initially calculated a Lomb-normalized periodogram over the
interval from the end of the outburst (around 1998 March 28) to \asmend.
The result was a peak with Lomb-normalized power 14.05 (equivalent to
3.9$\sigma$), at $383\pm30$~d.  We estimated the period
uncertainty by folding the ASM data on the detected period, and
calculating 1000 sets of statistically equivalent data sets with the same
time sampling, average profile and noise properties per bin as measured in
the original data set. The variance in the detection periods of the
simulated data sets is then a measure of the uncertainty in the actual
detection.

To obtain a more precise estimate of the quiescent X-ray periodicity, we
excluded those measurements with errors of $>0.5\ \cts$.  Around 13\% of
the 1-d averaged ASM measurements had errors above this threshold,
possibly due to source variability on time scales less than a day, but
more likely arising from low-significance detections due to solar
contamination or other sub-optimal observing conditions.  These
low-significance measurements appeared to introduce a bias to the period
measurements, and also tended to contribute disproportionately to the
estimated uncertainty in the period.  A Lomb-normalized periodogram
calculated on this subset of data following the outburst resulted in a
peak power of 24.04 (estimated significance $4.2\times10^{-8}$, equivalent to
$5.5\sigma$) at $392\pm14$~d.
To confirm the above results,
we used the phase-dispersion minimization technique of \cite{stell78}
and folded the ASM light curve on a grid of periods between 200 to 1000 d.
We found that the X-ray flux was most likely modulated with a period of
$404 \pm 13$ d.  Combining the results
from the latter two period-search techniques,
we determine the period of the X-ray oscillation
to be $400 \pm 20$~d with the epoch of maximum at MJD $51140 \pm 10$~d.
The mean profile of the post-outburst ASM measurements folded on the 400~d
period is also shown in Figure \ref{asm} (inset, bottom panel). The
profile was approximately sinusoidal with an amplitude of
$0.108 \pm 0.012$~ASM~$\cts$, or 31\% rms. With the periastron time from the
orbital fit used as a reference phase for the fold, we find that on
average the X-ray maximum occurs 0.1 in phase ($\sim40$~d) after
periastron.

For comparison, we also calculated an oversampled Lomb-normalized
periodogram of the entire ASM data set (including the 1997--98 outburst),
which revealed a significant periodicity at \pall~d (identical to that
measured following the outburst).  
The 1997--98 outburst was unusual; no similar events were
observed by the ASM in $\approx6.5$~yr of monitoring. Thus, it may have
occurred due to a mechanism distinct to some extent from the periodic
behavior of the quiescent emission. Furthermore, there is evidence that
the behaviour of the persistent emission prior to the outburst was
different than afterwards. Although only a little more than one 400-d
interval was covered before the outburst, the persistent flux was weakly
anticorrelated compared to the mean profile measured following the
outburst.  Applying the same error threshold ($\sigma<0.5\
\cts$) to the full data set resulted in an increase in both the peak
power and period, to 51.1 (estimated significance $1.2\times10^{-20}$ or
$>8\sigma$) at 385~d. 
The shorter measured periods using the full ASM data set may be attributed
to aliasing resulting from a delay between the nearest phase of X-ray
maximum for the persistent emission and the peak of the 1997--98 outburst.
If \srcx\ is similar to wind-accreting neutron stars in high-mass
binaries, we may expect large outbursts to correspond only approximately
with periastron phase \cite[e.g.][]{swr86}.  In fact, the peak of the 1997
outburst falls around MJD 50780, approximately $40$~d after the nearest
epoch of maximum flux according to the 400-d periodicity detected in
the post-1997 outburst ASM measurements, and approximately $70$ days after
periastron passage according to the orbital solution discussed in
\S\ref{sec:radvel} (Figure \ref{asm}).

\section{Discussion}


We have found evidence for significant variation in the radial
velocity measurements from \srco, which we fit with an eccentric orbit
with period $404\pm3$~d.  We have also found evidence for a
significant variation in the X-ray flux from \srcx.  Our best
estimate of the X-ray period is
$400 \pm 20$ d.  Hence, 
the periods measured from the radial velocity and ASM X-ray flux
variations are consistent.
It is compelling to suggest that these two periodicities arise from a
common source.
Naturally, it is also possible that the agreement between the two
periodicities is merely coincidence, and that \srco\ and \srcx\ are 
unrelated.  Estimating the probability of such a coincidence depends
upon the origin of the periodicities in each star, as well as the
intrinsic distributions of these phenomena.
In practice, it is not possible to make such an estimate with much
confidence.
However, 
it is clear that the likelihood of an unrelated red giant and X-ray
source being at nearly the same measured position at high galactic latitude,
and also both having a periodicity 
\centerline{\epsfxsize=8.5cm\epsfbox{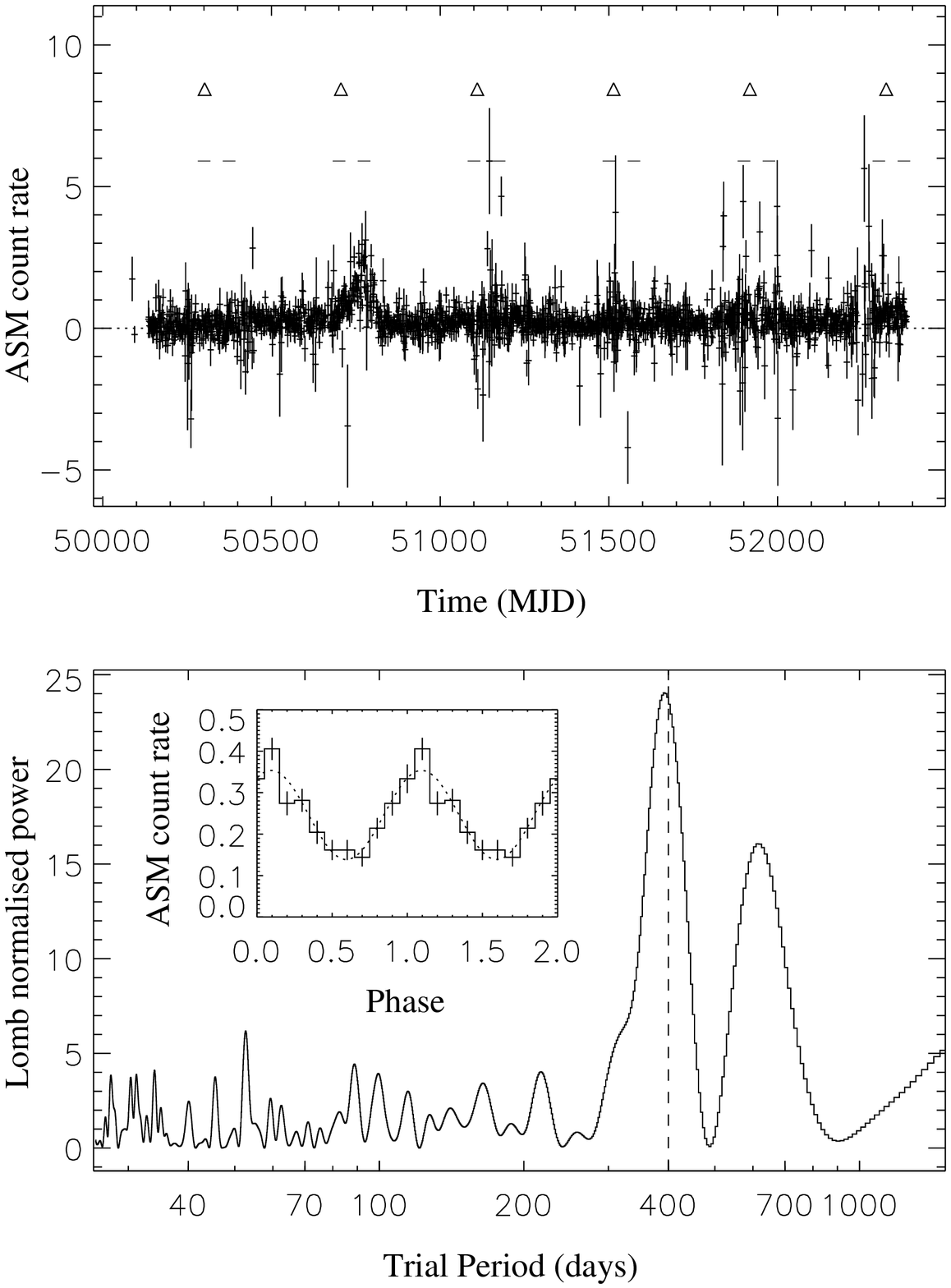}}
 \figcaption{\xte/ASM measurements of \srcx. The top
panel shows the 1~d averaged ASM measurements between 1996 January 5 and
\asmend\ (MJD 50087--\asmendmjd).  The horizontal lines (-- --) bracket
the estimated times of phase maximum for the best period estimate (400~d;
see \S\ref{sec:xte}).  The open triangles indicate the predicted times of
periastron passage for the eccentric orbital fit ($P_{\rm orb}=404$~d;
Table \ref{orbtbl}).  The bottom panel shows the periodogram calculated
from a subset of the ASM data after the end of the outburst (1998 March
28), where the measurement error was $<0.5\ \cts$. The dashed vertical
line shows the candidate 400~d period.  The inset shows the folded ASM
profile using the time of periastron from the orbital fit as the reference
phase.
 \label{asm} }
\bigskip
\noindent
that would be unusual for either type
of source separately, is extremely small.  The detection of a 400-d
modulation from both stars therefore supports the original suggestion that
HD 154791 is the optical counterpart of \srcx.


\subsection{Cause of the Correlation} \label{sec:causeofcor}

There are several plausible mechanisms for a correspondence between
the radial velocity and the persistent X-ray flux.
We consider two possibilities for the origin of the 404-d period in
the \srco\, radial velocity curve: orbital motion, and pulsational
variation of the M-giant.  Precise optical photometric monitoring
could distinguish between these two possibilities, since the maximum
brightness occurring at any phase besides that of maximum recessional
velocity would favour orbital motion.

If the radial velocity variations are due to orbital motion, the 400-d
X-ray flux modulation may arise from changes in the accretion rate
onto the neutron star as it follows an elliptical orbit around the
companion and moves through different nebular densities. Such
modulation is common in high-mass wind-accreting X-ray binaries
\cite[e.g.][]{swr86}.  Moreover, the times of periastron passage in
our best-fit orbital model are very well correlated with the times of
increased X-ray flux.  For a 1.4~$M_\odot$ neutron star and red giant
mass $\la 2.5\ M_\odot$, the 404~d period and low amplitude of the
radial velocity variations imply that the inclination is
$i\la2^\circ$.  Although the {\it a priori} probability of observing
such a system is $\la7\times10^{-4}$, such a low inclination could
explain the lack of X-ray pulsations or QPOs from the source (see 
\S\ref{gxcomp}).
If the 400-d variation is in fact due to orbital motion, the
longer-term variations in the radial velocities may be related to
irregular variations in the red giant \cite[]{pwh01}.

Alternatively, red-giant pulsations with $P=400$~d could 
modulate the stellar wind, and hence the accretion rate onto the neutron
star.  The IRAS fluxes of HD~154791 ($f_{12}=2.77$, $f_{25}=0.736$,
$f_{60}<0.4$, $f_{100}<1.0$, where $f_{12}$ is the flux density $f_\nu$ at
12 microns in Jy) indicate that is it not a Mira variable
\cite[]{kenyon88}.  However, non-Mira M-giants also pulsate.  Although
typical fundamental pulsation periods lie in the range 20--200~d,
variations with a period up to an order of magnitude larger are also seen
\cite[]{pwh01}.  If the 400-d variations are due to pulsation, one of the
slower trends observed in the radial velocity variations could be related
to orbital motion.  In this case, the possible range of orbital
inclinations becomes much more likely.  An orbital period of 2370~d gives
an inclination of 4--$6^\circ$ while for a period of 3700~d the likely
range is 5--$7^\circ$.  Also, the low accretion rate onto the neutron star
would be more understandable if the orbital period is longer than 400 d.
However, a longer orbital period alone cannot account for the discrepancy
in luminosity between 4U~1700+24 and GX~1+4.

Finally, the shape of the 400-d radial velocity variation argues
against a red-giant pulsation.  Pulsation light curves are typically
non-sinusoidal (in the sense that more time is spent with brightness
below the median value than above it), and similar behavior would be
expected for the radial velocity variations.

\subsection{Comparison with GX~1+4}
\label{gxcomp}

If the 400~d periodicity in the \srco\, radial velocities is due to
orbital motion, then the orbital periods of \srco/\srcx\ and GX~1+4 may be
quite similar (see Table \ref{comptable}). This similarity makes the
contrast between their other observed properties puzzling. The X-ray
luminosity of GX~1+4 appears to be two to three orders of magnitude
greater than that of \srcx.  In GX~1+4, there is clear evidence for an
accretion disk \cite[]{chak97:opt}, whereas in \srcx\, there is no
measurable UV continuum \cite[]{garcia83}.  GX~1+4 has a rich optical
emission-line spectrum, and \srco\, shows very little in the way of
emission lines.  Finally, GX~1+4 is a 2-minute X-ray pulsar, whereas no
coherent or quasi-period oscillations have been convincingly detected from
\srcx.

One explanation for the differing X-ray luminosities is a larger
mass-loss rate from the late-type giant in GX~1+4.  \cite{chak97:opt}
find that the M6 giant in GX~1+4 is probably near the tip
of the first-ascent red-giant branch.  \srco\, has a spectral type M2,
and so would be expected to be losing mass at a lower rate. 
The relative lack of optical/UV emission lines suggests that either
the ionized nebula in \srco/\srcx\ is substantially smaller and/or
less dense than in GX~1+4, or there are not enough UV photons to power
nebular line emission in \srcx/\srco.  Either of these conditions can
plausibly arise if the mass loss from the companion is much smaller.
Moreover,
if the binary separation is smaller in
GX~1+4 than in \srcx/\srco\ a larger fraction of the red giant wind could
be accreted.  The higher X-ray luminosity in GX~1+4
could also illuminate the red giant and increase the mass loss from
the red giant further, in a sense producing a feedback effect.  The
observed variability of the optical spectrum of V2116~Oph is evidence for
the significant role played by X-ray heating. On the other hand, in
\srcx/\srco\ the X-ray luminosity is perhaps low enough that very little
illumination of the red giant or enhanced mass loss is expected.
Even during the 1997 X-ray outburst, \cite{tomasella97} note that
the optical spectrum was unchanged.
Finally, the lack of pulsations or QPOs in \srcx\ remains puzzling.
\cite{garcia83} and \cite{mg01}
proposed a 900-s QPO, but this QPO was not confirmed in high-quality
\xte\/ data \cite[]{masetti01}.  High-precision optical photometry has also
revealed a lack of pulsations (down to a limit of 0.5\% fractional
variation), or rapid variations of any kind \cite[]{sbh01}.  This behavior
is again in contrast to GX~1+4, where \cite{jab97} find an oscillation in
the optical emission with an amplitude of from 0.4 to 4.5\%, plus
stochastic variations.  Given that the optical light from GX~1+4 has a
significant contribution from an accretion disk, the lack of rapid optical
variations from \srco\, provides evidence that any contribution from a
disk in \srcx\, is negligible in the optical regime as well as the UV
regime.  The lack of X-ray pulsations is expected if the 400-d radial
velocity variations are due to orbital motion.  As discussed in 
\S\ref{sec:causeofcor}, the low amplitude of the radial velocity
variations requires $i\sim 0$, in which case any magnetic hot spots on the
neutron star will be continuously in view (assuming the NS spin axis is
aligned with angular momentum axis of the binary).  For higher and {\it a
priori} more probable values of $i$, the non-detection of pulsations
may be explained if the
accretion flow is not appreciably disrupted by the magnetic field of the
neutron star above the surface \cite[e.g.][]{gl79a,gl79b}, and spreads the
accreting material more or less evenly over the neutron star surface.
Given the extremely low luminosity (and hence accretion rate) of this
source, standard calculations of ram and $B$-field pressure balance would
require that the surface magnetic field strength of the neutron star be
$\la10^6$~G, far below the canonical value for low-mass X-ray binaries or
even old neutron stars in recycled ms pulsars. A more likely scenario
consistent with larger $i$ is that the neutron star spin is very slow.


In conclusion, all scenarios for \srcx/\srco\, have some difficulties.
We have presented some options, but more observational constraints are
needed to select beween them.  Additional observations to determine
whether the 400-d radial velocity variations are due to orbital motion
or red-giant pulsation would also enable one to better interpret the
differences between \srcx/\srco\, and GX~1+4.



\acknowledgments

We are grateful to Dave Latham and Robert Stefanik for their assistance
with the data from the Wyeth telescope, and to Phil Uttley and Deepto
Chakrabarty for useful discussions.  This work was funded in part by NSF
grant INT-9902665 to J.L.S., and also the NASA Long Term Space
Astrophysics program under grant NAG 5-9184 (PI: Chakrabarty).  This
research has made use of data obtained through the High Energy
Astrophysics Science Archive Research Center Online Service, provided by
the NASA/Goddard Space Flight Center.

\clearpage




\begin{deluxetable}{lccl}
\tablewidth{0pt}
   \tablecaption{Comparison of the properties of the two known symbiotic
neutron-star binaries.  \label{comptable} }
  \tablehead{ \colhead{Parameter} & \colhead{4U~1700+24} & \colhead{GX 1+4} &
    \colhead{References} }
\startdata
  Distance (kpc) & $0.42\pm0.2$ & 3--6/12--15 \tablenotemark{a} & [1,2] \\
  $P_{\rm spin}$ (s) & ?   & $\approx100$--137.7 & [3,4] \\
  $P_{\rm orb}$ (d) & $404\pm3$? & 304? & [5,6,7] \\
  $L_{\rm X}$ (2--10~keV, $10^{36}\ \eps$) & (0.2--$10)\times10^{-3}$
    & $\sim10$/100\tablenotemark{b} & [1,2] \\
  $\dot{M}$ ($10^{16}\ {\rm g\,s^{-1}}$)\tablenotemark{c}
    & (0.1--$5)\times10^{-3}$ & 5/50\tablenotemark{b}\\
  Companion spectra & M2 III & M3--6 III & [1,2] \\
  $L_{\rm opt}/L_{\rm X}$ & 200 & 0.25 & [1] \\
\enddata
\tablenotetext{a}{Depending upon the evolutionary status of the mass donor, first
ascent red giant/beginning inital ascent of the asymptotic giant branch.}

\tablenotetext{b}{Depending upon the distance}

\tablenotetext{c}{Assuming a neutron star with $R_*=10$~km and $M=1.4M_\sun$.}

\tablerefs{
 1. \cite{masetti01}; 2. \cite{chak97:opt}; 3. \cite{damle88,leahy89}; 4.
\cite{jab02}; 5. this paper; 6.
\cite{cut86}; 7. \cite{per99} } 

\end{deluxetable}


\begin{table}
 \begin{center}
\caption{Orbital parameters for \srco\ \label{orbtbl} }
 \begin{tabular}{lc}
 \tableline
Radial velocity amplitude, $K$ (km$\,{\rm s}^{-1}$) & $0.75 \pm 0.12$ \\
Projected semimajor axis, $a_{\rm X} \sin i$ ($10^6$~km) & $4.2\pm0.7$ \\
Orbital period, $P_{\rm orb}$ (d) & $404\pm3$ \\
Epoch of periastron, $T_0$ (MJD) & $49090 \pm 80$ \\
Eccentricity, $e$ & $0.26\pm0.15$ \\
Longitude of periastron, $\omega$ (degrees) & $260 \pm 40$ \\
Mass function, $f_{\rm O}$ ($10^{-5}\ M_\odot$) & $1.8\pm0.9$ \\
Model fit $\chi^2$ & 95.96 (76 dof) \\
\tableline
 \end{tabular}
\end{center}
\end{table}


\end{document}